\documentclass[preprint,prd,aps,showpacs,showkeys,nofootinbib]{revtex4}
\usepackage{graphicx}
\begin{document}


\title{Neutrino mixing in the $\mu\nu$SSM}

\author{Hai-Bin Zhang$^{a,}$\footnote{email:hbzhang@mail.dlut.edu.cn}, Tai-Fu Feng$^{a,b,c,d}$, Li-Na Kou$^{a}$, Shu-Min Zhao$^{b,c,d}$}

\affiliation{$^a$Department of Physics, Dalian University of Technology,
Dalian, 116024, China\\
$^b$Department of Physics, Hebei University, Baoding, 071002, China\\
$^c$Institute of theoretical Physics, Chinese Academy of Sciences, Beijing 100190, China\\
$^d$The Key Laboratory of Mathematics-Mechanization (KLMM), Beijing 100190, China}

\begin{abstract}
Recently, several reactor oscillation experiments have successively measured a nonzero value for the neutrino mixing angle ${\theta _{13}}$, which is greater than 5 standard deviations. Within framework of the $\mu$ from $\nu$ Supersymmetric Standard Model ($\mu\nu$SSM), three tiny neutrino masses are generated at the tree level through TeV scale seesaw mechanism. In this work, we analyze the neutrino masses and mixing in the $\mu\nu$SSM with a ``top down'' method, assuming neutrino mass spectrum with normal ordering (NO) or inverted ordering (IO).
\end{abstract}


\maketitle

\section{Introduction\label{sec1}}
Recently the neutrino experiments develop quickly, which give the strong evidences for the massive neutrinos and their mixing. Three flavor neutrinos $\nu_{e,\mu,\tau}$ are mixed into three massive neutrinos $\nu_{1,2,3}$ during their flight, and the mixing is described by the Pontecorvo-Maki-Nakagawa-Sakata unitary matrix
$U_{_{PMNS}}$~\cite{oscillations,oscillations1}. Through the
several recent reactor oscillation experiments~\cite{theta13,theta13-1,theta13-2,theta13-3,theta13-4}, the neutrino mixing angle $\theta_{13}$ is now precisely known. The global fit of $\theta_{13}$ gives~\cite{Garcia}
\begin{eqnarray}
\sin^2\theta_{13}=0.023\pm 0.0023.
\label{neu-oscillations1}
\end{eqnarray}
The other experimental observations of the 3-neutrino oscillation parameters show that~\cite{PDG}
\begin{eqnarray}
&&\Delta m_{\odot}^2 =7.58_{-0.26}^{+0.22}\times 10^{-5} {\rm eV}^2,\nonumber\\
&&|\Delta m_{A}^2| =2.35_{-0.09}^{+0.12}\times 10^{-3} {\rm eV}^2,\nonumber\\
&&\sin^2\theta_{12} =0.306_{-0.015}^{+0.018},\qquad \sin^2\theta_{23}=0.42_{-0.03}^{+0.08}.
\label{neu-oscillations2}
\end{eqnarray}

Solving the $\mu$ problem~\cite{m-problem} of the Minimal Supersymmetric Standard Model (MSSM)~\cite{MSSM,MSSM1,MSSM2,MSSM3,MSSM4}, the ``$\mu$ from $\nu$ Supersymmetric Standard Model'' ($\mu\nu$SSM)~\cite{mnSSM,mnSSM1,mnSSM2,mnSSM3,mnSSM4,mnSSM5,mnSSM6,mnSSM7}, can generate three tiny neutrino masses at the tree level through TeV scale seesaw mechanism~\cite{mnSSM1,neutrino-mass,neu-mass1,neu-mass2,neu-mass3,neu-mass4,neu-mass5}. In this paper, we will analyse the three neutrino masses and mixing in the $\mu\nu$SSM with a ``top down'' method~\cite{top-down}, where the neutrino masses and the mixing angles are predicted from a given effective neutrino mass matrix. With the ``top down'' method, we can account for the experimental data on neutrino oscillations in the $\mu\nu$SSM, assuming neutrino mass spectrum with normal ordering (NO) or inverted ordering (IO).

This paper is organized as follows. In next section we outline the model by introducing its superpotential and neutrino sector. Through the ``top down'' method, the exact formulas of the neutrino masses and mixing angles for two possibilities
on the neutrino mass spectrum (NO and IO) are given in section~\ref{sec3}. In section~\ref{sec4} we give the numerical analysis, and we summarise in section~\ref{sec5}.

\section{the model and the neutrino sector\label{sec2}}
Besides the superfields of the MSSM, the $\mu\nu$SSM introduces three singlet right-handed neutrino superfields $\hat{\nu}_i^c$. In addition to the MSSM Yukawa couplings for charged leptons and quarks, the superpotential of the $\mu\nu$SSM contains Yukawa couplings for neutrinos, and two additional terms including the Higgs doublet superfields $\hat H_d$ and $\hat H_u$, and the right-handed neutrino superfields  $\hat{\nu}_i^c$,~\cite{mnSSM}
\begin{eqnarray}
&&W={\epsilon _{ab}}\left( {Y_{{u_{ij}}}}\hat H_u^b\hat Q_i^a\hat u_j^c + {Y_{{d_{ij}}}}\hat H_d^a\hat Q_i^b\hat d_j^c
+ {Y_{{e_{ij}}}}\hat H_d^a\hat L_i^b\hat e_j^c \right)  \nonumber\\
&&\hspace{0.95cm}
+ {\epsilon _{ab}}{Y_{{\nu _{ij}}}}\hat H_u^b\hat L_i^a\hat \nu _j^c -  {\epsilon _{ab}}{\lambda _i}\hat \nu _i^c\hat H_d^a\hat H_u^b + \frac{1}{3}{\kappa _{ijk}}\hat \nu _i^c\hat \nu _j^c\hat \nu _k^c ,
\end{eqnarray}
where $\hat H_d^T = \Big( {\hat H_d^0,\hat H_d^ - } \Big)$, $\hat H_u^T = \Big( {\hat H_u^ + ,\hat H_u^0} \Big)$, $\hat Q_i^T = \Big( {{{\hat u}_i},{{\hat d}_i}} \Big)$, $\hat L_i^T = \Big( {{{\hat \nu}_i},{{\hat e}_i}} \Big)$ are $SU(2)$ doublet superfields, and $\hat d_i^c$, $\hat u_i^c$ and $\hat e_i^c$ represent the singlet down-type quark, up-type quark and charged lepton superfields, respectively.  In addition, $Y$, $\lambda$ and $\kappa$ are dimensionless matrices, a vector and a totally symmetric tensor.  $a,b=1,2$ are SU(2) indices with antisymmetric tensor $\epsilon_{12}=-\epsilon_{21}=1$, and $i,j,k=1,2,3$ are generation indices. The summation convention is implied on repeated indices in this paper.

In the superpotential, if the scalar potential is such that nonzero  vacuum expectation values (VEVs) of the scalar components ($\tilde \nu _i^c$) of the right-handed neutrino superfields $\hat{\nu}_i^c$ are induced, the effective bilinear terms $\epsilon _{ab} \varepsilon_i \hat H_u^b\hat L_i^a$ and $\epsilon _{ab} \mu \hat H_d^a\hat H_u^b$ are generated, with $\varepsilon_i= Y_{\nu _{ij}} \left\langle {\tilde \nu _j^c} \right\rangle$ and $\mu  = {\lambda _i}\left\langle {\tilde \nu _i^c} \right\rangle$,  once the electroweak symmetry is broken (EWSB). And the last term can generate the effective neutrino Majorana masses at the electroweak scale.

In addition, the last two terms in the superpotential explicitly violate lepton number and R-parity. In supersymmetric (SUSY) extensions of the Standard Model (SM), the R-parity of a particle is defined as $R = (-1)^{L+3B+2S}$~\cite{MSSM} and can be violated if either the baryon number ($B$) or lepton number ($L$) is not conserved, where $S$ denotes the spin of concerned component field. Note that $R=+1$ for particles and $-1$ for superparticles.

The phenomenology of the model where R-parity is broken differs substantially from that
of a model where R-parity is conserved. R-parity breaking implies that the Lightest Supersymmetric Particle (LSP) is no longer stable. In this context, the neutralino or the sneutrino are no longer candidates for the dark matter. However, other SUSY particles such as the gravitino or the axino can still be used as candidates.~\cite{mnSSM1}

Once EWSB, the neutral scalars develop in general the following VEVs:
\begin{eqnarray}
\langle H_d^0 \rangle = \upsilon_d \:, \qquad \langle H_u^0 \rangle = \upsilon_u \:, \qquad
\langle \tilde \nu_i \rangle = \upsilon_{\nu_i} \:, \qquad \langle \tilde \nu_i^c \rangle = \upsilon_{\nu_i^c} \:.
\end{eqnarray}
Then, the mass squared of charged gauge boson is given as
\begin{eqnarray}
m_W^2= \frac{e^2}{2 s_{_W}^2} \Big(\upsilon_u^2 + \upsilon_d^2 + \upsilon_{\nu_i}\upsilon_{\nu_i} \Big),
\end{eqnarray}
and one can define
\begin{eqnarray}
\tan\beta=\frac{\upsilon_u}{\sqrt{\upsilon_d^2+\upsilon_{\nu_i}\upsilon_{\nu_i}}}.
\end{eqnarray}
Here $e$ is the electromagnetic coupling constant, $s_{_W}=\sin\theta_{_W}$ with the Weinberg angle $\theta_{_W}$, respectively.

Recently, the ATLAS~\cite{ATLAS} and CMS~\cite{CMS} have reported significant events which are interpreted to be related to the neutral SM-like Higgs boson with mass around 125~GeV. This fact constrains parameter space of the $\mu\nu$SSM stringently. Through the Ref.~\cite{mnSSM6}, we investigate the radiative correction to the SM-like Higgs boson mass. Besides the superfields of the MSSM, the $\mu\nu$SSM introduces right-handed neutrino superfields. Nevertheless the loop effects of right-handed neutrino/sneutrino on the SM-like Higgs boson mass can be neglected, due to small neutrino Yukawa couplings $Y_{\nu_i} \sim \mathcal{O}(10^{-7})$ and left-handed neutrino superfield VEVs $\upsilon_{\nu_i} \sim \mathcal{O}(10^{-4}{\rm{GeV}})$.

In the $\mu\nu$SSM, neutrinos are mixed with the neutralinos~\cite{mnSSM,mnSSM1}. In the weak interaction basis defined by ${\chi '^{ 0 T}} = \left( {{{\tilde B}^ 0 },{{\tilde W}^ 0 },{{\tilde H}_d}{\rm{,}}{{\tilde H}_u},{\nu_{R_i}},{\nu_{L{_i}}}} \right)$, one can obtain the neutral fermion mass terms in the Lagrangian:
\begin{eqnarray}
\mathcal{L}_{neutral}^{mass} = - \frac{1}{2}{\chi '^{ 0 T}}{M_n}{\chi '^ 0 } + {\rm{H.c.}}  ,
\end{eqnarray}
with
\begin{eqnarray}
{M_n} = \left( {\begin{array}{*{20}{c}}
   M & {{m^T}}  \\
   m & {{0_{3 \times 3}}}  \\
\end{array}} \right).
\end{eqnarray}
Here, the submatrix $m$ is neutralino-neutrino mixing
\begin{eqnarray}
m = \left( {\begin{array}{*{20}{c}}
   {  -\frac{g_1}{\sqrt 2 }\upsilon_{{\nu _1}}} & { \frac{g_2}{\sqrt 2 }\upsilon_{{\nu _1}}} & 0 & {{Y_{{\nu _{1i}}}}{\upsilon_{\nu _i^c}}} & {{Y_{{\nu _{11}}}}{\upsilon_u}} & {{Y_{{\nu _{12}}}}{\upsilon_u}} & {{Y_{{\nu _{13}}}}{\upsilon_u}}  \\
   {  -\frac{g_1}{\sqrt 2 }\upsilon_{{\nu _2}}} & { \frac{g_2}{\sqrt 2 }\upsilon_{{\nu _2}}} & 0 & {{Y_{{\nu _{2i}}}}{\upsilon_{\nu _i^c}}} & {{Y_{{\nu _{21}}}}{\upsilon_u}} & {{Y_{{\nu _{22}}}}{\upsilon_u}} & {{Y_{{\nu _{23}}}}{\upsilon_u}}  \\
   {  -\frac{g_1}{\sqrt 2 }\upsilon_{{\nu _3}}} & { \frac{g_2}{\sqrt 2 }\upsilon_{{\nu _3}}} & 0 & {{Y_{{\nu _{3i}}}}{\upsilon_{\nu _i^c}}} & {{Y_{{\nu _{31}}}}{\upsilon_u}} & {{Y_{{\nu _{32}}}}{\upsilon_u}} & {{Y_{{\nu _{33}}}}{\upsilon_u}}  \\
\end{array}} \right).
\end{eqnarray}
And the submatrix $M$ is neutralino mass matrix including right-handed neutrinos
\begin{eqnarray}
&&M = \left( {\begin{array}{*{20}{c}}
   {{M_1}} & 0 & {\frac{-g_1}{{\sqrt 2 }}{\upsilon _d}} & {\frac{g_1}{{\sqrt 2 }}{\upsilon _u}} & 0 & 0 & 0  \\
   0 & {{M_2}} & {\frac{g_2}{{\sqrt 2 }}{\upsilon _d}} & {\frac{-g_2}{{\sqrt 2 }}{\upsilon _u}} & 0 & 0 & 0  \\
   {\frac{-g_1}{{\sqrt 2 }}{\upsilon _d}} & {\frac{g_2}{{\sqrt 2 }}{\upsilon _d}} & 0 & {-{\lambda _i}{\upsilon _{\nu _i^c}}} & { - {\lambda _1}{\upsilon _u}} & { - {\lambda _2}{\upsilon _u}} & { - {\lambda _3}{\upsilon _u}}  \\
   {\frac{g_1}{{\sqrt 2 }}{\upsilon _u}} & {\frac{-g_2}{{\sqrt 2 }}{\upsilon _u}} & {-{\lambda _i}{\upsilon _{\nu _i^c}}} & 0 & {y_1} & {y_2} & { y_3}  \\
   0 & 0 & { - {\lambda _1}{\upsilon _u}} & { y_1} & {2{\kappa _{11j}}{\upsilon _{\nu _j^c}}} & {2{\kappa _{12j}}{\upsilon _{\nu _j^c}}} & {2{\kappa _{13j}}{\upsilon _{\nu _j^c}}}  \\
   0 & 0 & { - {\lambda _2}{\upsilon _u}} & { y_2} & {2{\kappa _{21j}}{\upsilon _{\nu _j^c}}} & {2{\kappa _{22j}}{\upsilon _{\nu _j^c}}} & {2{\kappa _{23j}}{\upsilon _{\nu _j^c}}}  \\
   0 & 0 & { - {\lambda _3}{\upsilon _u}} & { y_3} & {2{\kappa _{31j}}{\upsilon _{\nu _j^c}}} & {2{\kappa _{32j}}{\upsilon _{\nu _j^c}}} & {2{\kappa _{33j}}{\upsilon _{\nu _j^c}}}  \\
\end{array}} \right),
\end{eqnarray}
where $y_i=- {\lambda _i}{\upsilon _d}+ {{Y_{{\nu _{ji}}}}{\upsilon _{{\nu _j}}} }$.

Via the seesaw mechanism, the effective light neutrino mass matrix is in general given as~\cite{mnSSM1,neutrino-mass}
\begin{eqnarray}
{m_{eff}} =  -  m.{M^{ - 1}} .{m^T}.
\end{eqnarray}
Diagonalized the effective neutrino mass matrix ${m_{eff}}$, we can obtain three light neutrino masses.

\section{Neutrino masses and mixing\label{sec3}}
With the ``top down'' method~\cite{top-down}, we give the exact formulas for the neutrino masses and mixing angles from the given effective neutrino mass matrix ${m_{eff}}$ in this section. For the given effective neutrino mass matrix ${m_{eff}}$, we diagonalize the hermitian matrix
\begin{eqnarray}
{\cal H}={m_{eff}}^{\dagger}m_{eff}.
\end{eqnarray}

The eigenvalues of the $3\times3$  effective neutrino mass squared matrix ${\cal H}$ are given as
\begin{eqnarray}
&&m_1^2={a\over3}-{1\over3}p(\cos\phi+\sqrt{3}\sin\phi),
\nonumber\\
&&m_2^2={a\over3}-{1\over3}p(\cos\phi-\sqrt{3}\sin\phi),
\nonumber\\
&&m_3^2={a\over3}+{2\over3}p\cos\phi.
\end{eqnarray}
To formulate the expressions in a concise form, one can define the notations
\begin{eqnarray}
&&p=\sqrt{a^2-3b},\qquad \phi={1\over3}\arccos({1\over p^3}(a^3-{9\over2}ab+{27\over2}c))
\end{eqnarray}
with
\begin{eqnarray}
&&a={\rm Tr}({\cal H}),\nonumber\\
&&b={\cal H}_{11}{\cal H}_{22}+{\cal H}_{11}{\cal H}_{33}+{\cal H}_{22}{\cal H}_{33}
-{\cal H}_{12}^2-{\cal H}_{13}^2-{\cal H}_{23}^2,\nonumber\\
&&c={\rm Det}({\cal H}).
\end{eqnarray}
To account for the experimental data on neutrino oscillations, we check that $m_1^2<m_2^2<m_3^2$.
In the case of 3-neutrino mixing, we have two possibilities
on the neutrino mass spectrum~\cite{PDG}
\begin{itemize}
\item (i) spectrum with normal ordering (NO):
\begin{eqnarray}
&&m_{\nu_1}<m_{\nu_2}<m_{\nu_3}, \nonumber\\
&&m_{\nu_1}^2=m_1^2,\quad m_{\nu_2}^2=m_2^2,\quad m_{\nu_3}^2=m_3^2, \nonumber\\
&&\Delta m_{\odot}^2 = m_{\nu_2}^2-m_{\nu_1}^2 ={2\over \sqrt{3}}p\sin\phi>0,\nonumber\\
&&\Delta m_{A}^2 = m_{\nu_3}^2-m_{\nu_1}^2 =p(\cos\phi+{1\over\sqrt{3}}\sin\phi)>0;
\end{eqnarray}

\item (ii) spectrum with inverted ordering (IO):
\begin{eqnarray}
&&m_{\nu_3}<m_{\nu_1}<m_{\nu_2}, \nonumber\\
&&m_{\nu_3}^2=m_1^2,\quad m_{\nu_1}^2=m_2^2,\quad m_{\nu_2}^2=m_3^2, \nonumber\\
&&\Delta m_{\odot}^2 = m_{\nu_2}^2-m_{\nu_1}^2 =p(\cos\phi-{1\over\sqrt{3}}\sin\phi)>0, \nonumber\\
&&\Delta m_{A}^2 = m_{\nu_3}^2-m_{\nu_2}^2 =-p(\cos\phi+{1\over\sqrt{3}}\sin\phi)<0.
\end{eqnarray}
\end{itemize}

The normalized eigenvectors for the mass squared matrix ${\cal H}$ are given as
\begin{eqnarray}
&&\left(\begin{array}{c}\Big(U_\nu\Big)_{11}\\
\Big(U_\nu\Big)_{21}\\\Big(U_\nu\Big)_{31}
\end{array}\right)={1\over\sqrt{|X_1|^2+|Y_1|^2+|Z_1|^2}}\left(\begin{array}{c}
X_1\\Y_1\\Z_1\end{array}\right),
\nonumber\\
&&\left(\begin{array}{c}\Big(U_\nu\Big)_{12}\\
\Big(U_\nu\Big)_{22}\\\Big(U_\nu\Big)_{32}
\end{array}\right)={1\over\sqrt{|X_2|^2+|Y_2|^2+|Z_2|^2}}\left(\begin{array}{c}
X_2\\Y_2\\Z_2\end{array}\right),
\nonumber\\
&&\left(\begin{array}{c}\Big(U_\nu\Big)_{13}\\
\Big(U_\nu\Big)_{23}\\\Big(U_\nu\Big)_{33}
\end{array}\right)={1\over\sqrt{|X_3|^2+|Y_3|^2+|Z_3|^2}}\left(\begin{array}{c}
X_3\\Y_3\\Z_3\end{array}\right),
\end{eqnarray}
with
\begin{eqnarray}
&&X_1=({\cal H}_{22}-m_{{\nu_1}}^2)({\cal H}_{33}-m_{{\nu_1}}^2)-{\cal H}_{23}^2, \nonumber\\
&&Y_1={\cal H}_{13}{\cal H}_{23}-{\cal H}_{12}({\cal H}_{33}-m_{{\nu_1}}^2), \nonumber\\
&&Z_1={\cal H}_{12}{\cal H}_{23}-{\cal H}_{13}({\cal H}_{22}-m_{{\nu_1}}^2),
\nonumber\\
&&X_2={\cal H}_{13}{\cal H}_{23}-{\cal H}_{12}\Big({\cal H}_{33}-m_{{\nu_2}}^2\Big),
\nonumber\\
&&Y_2=({\cal H}_{11}-m_{{\nu_2}}^2)({\cal H}_{33}-m_{{\nu_2}}^2)-{\cal H}_{13}^2,
\nonumber\\
&&Z_2={\cal H}_{12}{\cal H}_{13}-{\cal H}_{23}\Big({\cal H}_{11}-m_{{\nu_2}}^2\Big),
\nonumber\\
&&X_3={\cal H}_{12}{\cal H}_{23}-{\cal H}_{13}\Big({\cal H}_{22}-m_{{\nu_3}}^2\Big),
\nonumber\\
&&Y_3={\cal H}_{12}{\cal H}_{13}-{\cal H}_{23}\Big({\cal H}_{11}-m_{{\nu_3}}^2\Big),
\nonumber\\
&&Z_3=({\cal H}_{11}-m_{{\nu_3}}^2)({\cal H}_{22}-m_{{\nu_3}}^2)-{\cal H}_{12}^2.
\end{eqnarray}
Correspondingly, the mixing angles among three tiny neutrinos are determined by
\begin{eqnarray}
&&\sin\theta_{13}=\Big|\Big(U_\nu\Big)_{13}\Big|,\qquad\qquad\:
\cos\theta_{13}=\sqrt{1-\Big|\Big(U_\nu\Big)_{13}\Big|^2}, \nonumber\\
&&\sin\theta_{23}={\Big|\Big(U_\nu\Big)_{23}\Big|\over\sqrt{1-\Big|\Big(U_\nu\Big)_{13}\Big|^2}},\quad
\cos\theta_{23}={\Big|\Big(U_\nu\Big)_{33}\Big|\over\sqrt{1-\Big|\Big(U_\nu\Big)_{13}\Big|^2}}, \nonumber\\
&&\sin\theta_{12}={\Big|\Big(U_\nu\Big)_{12}\Big|\over\sqrt{1-\Big|\Big(U_\nu\Big)_{13}\Big|^2}},\quad
\cos\theta_{12}={\Big|\Big(U_\nu\Big)_{11}\Big|\over\sqrt{1-\Big|\Big(U_\nu\Big)_{13}\Big|^2}}.
\end{eqnarray}

\section{Numerical analysis\label{sec4}}
In this numerical analysis, we can adopt the minimal flavor violation (MFV) assumption
\begin{eqnarray}
{\kappa _{ijk}} = \kappa {\delta _{ij}}{\delta _{jk}}, \qquad
\lambda _i = \lambda, \qquad \upsilon_{\nu_i^c}=\upsilon_{\nu^c},\qquad {Y_{{\nu _{ij}}}} = {Y_{{\nu _i}}}{\delta _{ij}}.
\end{eqnarray}
Then, the effective light neutrino mass matrix $m_{eff}$ can approximate as~\cite{neutrino-mass}
\begin{eqnarray}
{m_{ef{f_{ij}}}} \approx \frac{{2A{\upsilon_{\nu^c}}}}{{3\Delta }}{b_i}{b_j} + \frac{{1 - 3{\delta _{ij}}}}{{6\kappa {\upsilon_{\nu^c}}}}{a_i}{a_j},
\end{eqnarray}
where
\begin{eqnarray}
&&\Delta  = {\lambda ^2}{(\upsilon_d^2 + \upsilon_u^2)}^2 + 4\lambda \kappa {\upsilon_{\nu^c}^2}{\upsilon_d}{\upsilon_u} - 12{\lambda ^2}{\upsilon_{\nu^c}}AB,\nonumber\\
&& A = \kappa {\upsilon_{\nu^c}^2} + \lambda {\upsilon_d}{\upsilon_u},\quad \frac{1}{B}= \frac{e^2}{c_{_W}^2{M_1}} + \frac{e^2}{s_{_W}^2{M_2}} , \nonumber\\
&&{a_i} = {Y_{{\nu _i}}}{\upsilon_u}\:, \qquad\qquad\;\: {b_i} = {Y_{{\nu _i}}}{\upsilon_d} + 3\lambda \upsilon_{\nu_i}.
\end{eqnarray}

Through the discussion of the parameter space in Refs.~\cite{mnSSM5,mnSSM6,mnSSM7}, we choose the relevant parameters as default in our numerical calculation:
\begin{eqnarray}
&&\tan \beta=1.4,\qquad \kappa=0.01,\qquad  \lambda=0.1,  \nonumber\\
&&\upsilon_{\nu^c}=800\:{\rm{GeV}},\qquad  M_2=2M_1=3\:{\rm{TeV}}.
\end{eqnarray}

In order to fit the experimental data on neutrino mass squared differences and mixing angles in Eq.(\ref{neu-oscillations1}) and  Eq.(\ref{neu-oscillations2}) with two possibilities of the neutrino mass spectrum, we choose the VEVs of left-handed neutrino superfields and the Yukawa couplings of right-handed neutrinos respectively as
\begin{itemize}
\item (i) For the NO spectrum:
\begin{eqnarray}
&&\upsilon_{\nu_1}=3.53\times10^{-4}\:{\rm GeV},\;\;
\upsilon_{\nu_2}=0.33\times10^{-4}\:{\rm GeV},\;\;
\upsilon_{\nu_3}=1.81\times10^{-4}\:{\rm GeV},
\nonumber\\
&&Y_{\nu_1}=0.931\times10^{-7},\qquad\:
Y_{\nu_2}=2.039\times10^{-7},\qquad\:
Y_{\nu_3}=2.010\times10^{-7},
\end{eqnarray}
and issue the theoretical predictions on light neutrino masses and mixing angles as
\begin{eqnarray}
&&m_{{\nu_1}}^2=5.180\times10^{-4}\:{\rm eV^2},\;\;m_{{\nu_2}}^2=5.943\times10^{-4}\:{\rm eV^2},\;\;
m_{{\nu_3}}^2=2.866\times10^{-3}\;{\rm eV^2},
\nonumber\\
&&\sin^2\theta_{{12}}=0.3058,\qquad\:\, \sin^2\theta_{{23}}=0.4202,\qquad\:\, \sin^2\theta_{{13}}=0.0232.
\end{eqnarray}

\item (ii) For the IO spectrum:
\begin{eqnarray}
&&\upsilon_{\nu_1}=4.25\times10^{-4}\:{\rm GeV},\;\;
\upsilon_{\nu_2}=2.72\times10^{-4}\:{\rm GeV},\;\;
\upsilon_{\nu_3}=2.65\times10^{-4}\:{\rm GeV},
\nonumber\\
&&Y_{\nu_1}=2.206\times10^{-7},\qquad\:
Y_{\nu_2}=2.382\times10^{-7},\qquad\:
Y_{\nu_3}=2.323\times10^{-7},
\end{eqnarray}
and issue the theoretical predictions on light neutrino masses and mixing angles as
\begin{eqnarray}
&&m_{{\nu_3}}^2=2.495\times10^{-3}\:{\rm eV^2},\;\;m_{{\nu_1}}^2=4.784\times10^{-3}\:{\rm eV^2},\;\;
m_{{\nu_2}}^2=4.860\times10^{-3}\;{\rm eV^2},
\nonumber\\
&&\sin^2\theta_{{12}}=0.3098,\qquad\:\, \sin^2\theta_{{23}}=0.4229,\qquad\:\, \sin^2\theta_{{13}}=0.0228.
\end{eqnarray}
\end{itemize}

\begin{figure}[htbp]
\setlength{\unitlength}{1mm}
\centering
\begin{minipage}[c]{0.5\textwidth}
\includegraphics[width=3.2in]{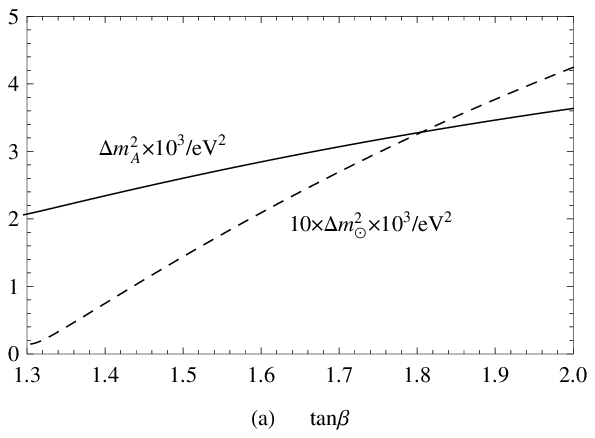}
\end{minipage}%
\begin{minipage}[c]{0.5\textwidth}
\includegraphics[width=3.2in]{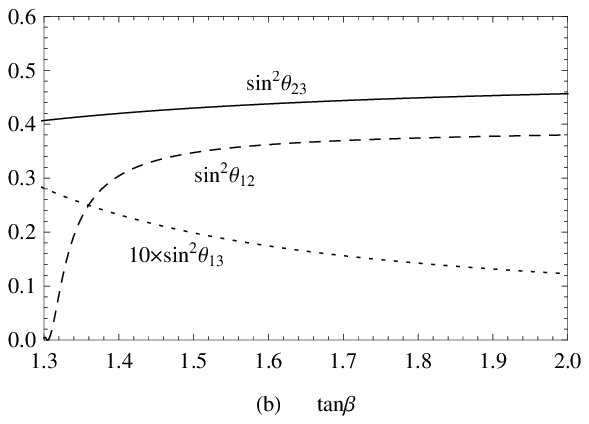}
\end{minipage}
\caption[]{Assuming neutrino mass spectrum with NO, we plot the neutrino mass squared differences and mixing angles versus  $\tan\beta$, where (a) $\Delta m_{A}^2$ (solid line) and $\Delta m_{\odot}^2$ (dashed line) versus $\tan\beta$, and (b) $\sin^2\theta_{{23}}$ (solid line), $\sin^2\theta_{{12}}$ (dashed line) and $\sin^2\theta_{{13}}$ (dotted line) versus $\tan\beta$, respectively.}
\label{fig-tb-N}
\end{figure}

\begin{figure}[htbp]
\setlength{\unitlength}{1mm}
\centering
\begin{minipage}[c]{0.5\textwidth}
\includegraphics[width=3.2in]{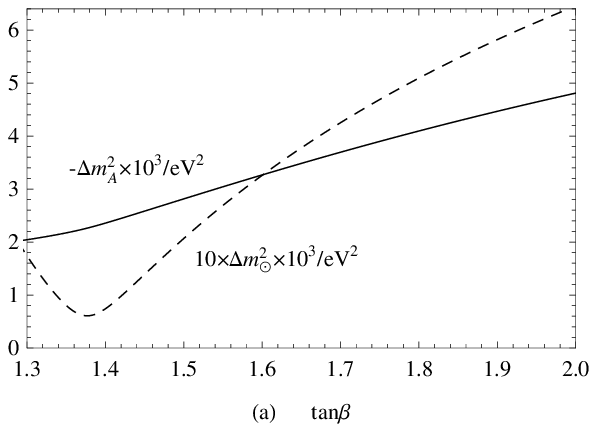}
\end{minipage}%
\begin{minipage}[c]{0.5\textwidth}
\includegraphics[width=3.2in]{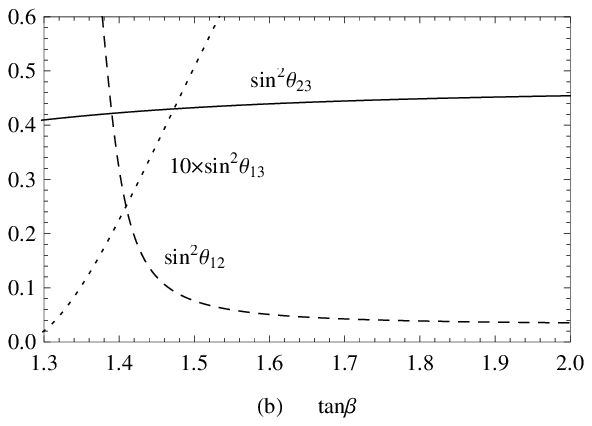}
\end{minipage}
\caption[]{Assuming neutrino mass spectrum with IO, we plot the neutrino mass squared differences and mixing angles versus  $\tan\beta$, where (a) $\Delta m_{A}^2$ (solid line) and $\Delta m_{\odot}^2$ (dashed line) versus $\tan\beta$, and (b) $\sin^2\theta_{{23}}$ (solid line), $\sin^2\theta_{{12}}$ (dashed line) and $\sin^2\theta_{{13}}$ (dotted line) versus $\tan\beta$, respectively.}
\label{fig-tb-I}
\end{figure}

Assuming neutrino mass spectrum with NO, we plot the neutrino mass squared differences varying with $\tan\beta$ in Fig.\ref{fig-tb-N}(a), where the solid line denotes $\Delta m_{A}^2$ versus $\tan\beta$ and the dashed line denotes $\Delta m_{\odot}^2$
versus $\tan\beta$. Along with increasing of $\tan\beta$, the theoretical prediction on $\Delta m_{A}^2$ increases gently, and the theoretical prediction on $\Delta m_{\odot}^2$ increases steeply. Under the same choice on the parameter space, we also draw the neutrino mixing angles varying with $\tan\beta$ in Fig.\ref{fig-tb-N}(b), where the solid line denotes $\sin^2\theta_{{23}}$ versus $\tan\beta$, the dashed line denotes $\sin^2\theta_{{12}}$ versus $\tan\beta$, and the dotted line denotes $\sin^2\theta_{{13}}$ versus $\tan\beta$, respectively. It shows that the mixing angle $\theta_{{12}}$ depends on $\tan\beta$ acutely as $\tan\beta\leq1.5$, and other mixing angles $\theta_{{23}}$ and $\theta_{{13}}$
vary with $\tan\beta$ mildly.

When the neutrino mass spectrum is IO, we also depict the neutrino mass squared differences and mixing angles varying with $\tan\beta$ in Fig.\ref{fig-tb-I}, where (a) $\Delta m_{A}^2$ (solid line) and $\Delta m_{\odot}^2$ (dashed line) versus $\tan\beta$, and (b) $\sin^2\theta_{{23}}$ (solid line), $\sin^2\theta_{{12}}$ (dashed line) and $\sin^2\theta_{{13}}$ (dotted line) versus $\tan\beta$, respectively. With increasing of $\tan\beta$, the theoretical prediction on $\Delta m_{A}^2$  and $\Delta m_{\odot}^2$ all increase, when $\tan\beta\geq1.4$. In Fig.\ref{fig-tb-I}(b), one can find that the mixing angles $\theta_{{12}}$ and $\theta_{{13}}$ depend on $\tan\beta$ acutely as $\tan\beta\leq1.6$, and other mixing angle $\theta_{{23}}$ vary with $\tan\beta$ gently.

\section{Summary\label{sec5}}
Through TeV scale seesaw mechanism, the $\mu\nu$SSM generates the effective neutrino mass matrix ${m_{eff}}$. From the given matrix ${m_{eff}}$, we give the exact formulas for the neutrino masses and mixing angles with the ``top down'' method. Under some assumption of the parameter space, the theoretical predictions on the neutrino mass squared differences and mixing angles can account for the experimental data on neutrino oscillations in Eq.(\ref{neu-oscillations1}) and  Eq.(\ref{neu-oscillations2}), assuming neutrino mass spectrum with normal ordering (NO) or inverted ordering (IO).

\begin{acknowledgments}
\indent\indent
The work has been supported by the National Natural Science Foundation of China (NNSFC)
with Grant No. 11275036, No. 11047002, the open project of State
Key Laboratory of Mathematics-Mechanization with Grant No. Y3KF311CJ1, the Natural
Science Foundation of Hebei province with Grant No. A2013201277, and Natural Science Fund of Hebei University with Grant No. 2011JQ05, No. 2012-242.
\end{acknowledgments}

\end{document}